On speed of a photon in a dispersing medium

Ogluzdin Valeriy E.

Abstract

Based on the author of the experimental results and their treatment, a model that demonstrates that the real dispersive medium in spectral regions where the refractive index greater than unity, the velocity of propagation of photons in this environment corresponds to the phase velocity of light. These portions of the spectrum (for example, atomic potassium vapor or other alkali metal) are from low-frequency side of the main lines of the doublet and are available for research using tunable lasers.
In the calculations took into account the fact that the atomic vapor of alkali metal used as a dispersing medium, in accordance with the theory of dispersion in the above sections of the spectrum, on the periphery of the cross section of laser beam with Gaussian intensity distribution across the beam - in the shadows refractive index $n(v)$ is greater than unity, and the axis of the beam, where the radiation intensity is maximum - for radiation following the leading edge of the light pulse, the refractive index is unity, which ensures smooth on the beam axis, without any delay "superluminal "propagation of light with velocity $c$. This fact causes the appearance of the shadow cone of Vavilov - Cherenkov radiation, indicating a slow propagation of photons from the source of the cone with phase velocity $c/n(v)$.

Valeriy E. Ogluzdin
Sovjet Union, 119991,Moscow , 38 Vavilova st.
Prokhorovs General Physics Institute.
Контактный тел. (499) 503 81 98
Факс: (499)135 81 91
ogluzdin@kapella.gpi.ru

# On speed of a photon in a dispersing medium

Valeriy E. Ogluzdin

Sovjet Union, 119991, Moscow , 38 Vavilova st.

Prokhorovs General Physics Institute.

## О СКОРОСТИ ФОТОНОВ В ДИСПЕРГИРУЮЩЕЙ СРЕДЕ

В.Е.Оглуздин

ИОФ РАН, Москва, 119991, ул. Вавилова д. 38

ogluzdin@kapella.gpi.ru

**Аннотация**

На основе полученных автором экспериментальных результатов и их обработки предлагается модель, свидетельствующая о том, что в реальной диспергирующей среде на участках спектра, где показатель преломления среды больше единицы, скорость распространения фотонов в такой среде соответствует величине фазовой скорости света. Такие участки спектра (на примере атомарных паров калия или другого щелочного металла), находятся с низкочастотной стороны линий главного дублета и оказываются доступными для исследований с помощью перестраиваемых по частоте лазеров.

При расчетах принималось во внимание то обстоятельство, что в атомарных парах щелочного металла, используемых в качестве диспергирующей среды, в соответствии с теорией дисперсии, на вышеуказанных участках спектра, на периферии сечения (в области тени) пучка лазерного излучения с гауссовым распределением интенсивности по сечению пучка, показатель преломления $n(v)$ больше единицы, а на оси светового пучка, там, где интенсивность излучения максимальна - для излучения, следующего за передним фронтом светового импульса, показатель преломления стремится к единице, что обеспечивает на оси пучка беспрепятственное, без какого-либо замедления "сверхсветовое" распространение светового излучения со скоростью $c$. Данное обстоятельство обуславливает появление в области тени конуса черенковского излучения, свидетельствующего о замедленном распространении фотонов по образующим конуса с фазовой скоростью $c / n(v)$.

### 1. Введение.

На основе полученных автором экспериментальных результатов [1,2] предлагается модель, свидетельствующая о том, что в реальной диспергирующей среде на участках спектра, где показатель преломления среды больше единицы, скорость распространения фотонов в такой среде соответствует величине фазовой скорости света $V = c / n(v)$. Такие участки спектра в атомарных парах калия (или другого щелочного металла), находятся с низкочастотной стороны линий главного дублета и оказываются доступными для исследований с помощью перестраиваемых по частоте лазеров.

### 2. Основная часть.

В работах [1,2] сообщалось о наблюдении с помощью перестраиваемого по частоте параметрического генератора света на выходе кюветы с атомарными парами калия конусного излучения, причем наблюдаемый в эксперименте угол раствора конуса соответствовал величине, получаемой в результате расчета, основанного на предположении о возможности использования для интерпретации экспериментальных данных модели излучения Вавилова-Черенкова [3].

Численные расчеты, выполненные в [1] без каких-либо подгоночных параметров на основе известных из классической физики соотношений для показателя преломления $n(v)$ [4], и для угла раствора черенковского конуса, подтверждали результаты эксперимента в области спектра, где показатель преломления среды больше единицы $n(v) > 1$. Отметим, что роль источника черенковского излучения в [1] отводилась наводимой и распространяющейся со "сверхсветовой" скоростью в среде двухуровневых атомов нелинейной поляризации, связанной с шестифотонным параметрическим рассеянием света (ШПР) [1].

Из представленных в [1] результатов эксперимента следует тот замечательный факт, что движение фотонов вдоль образующих черенковского конуса осуществляется с учетом величины показателя преломления $n(\nu_{as})$ со скоростью $V_{as} = c/n(\nu_{as})$, что явным образом соответствует фазовой скорости светового излучения. В рассмотренном случае частота $\nu_{as}$ смещена в антистоксову область спектра относительно частоты излучения накачки $\nu$. Ниже, приведено соотношение (3), согласно которому может быть рассчитана частота $\nu_{as}$.

Подобная ситуация складывается также в случае, если на оси пучка с гауссовым распределением интенсивности по его сечению помимо нелинейной поляризации, соответствующей ШПР, которое в области частот $\nu < \nu_{01}$ способствует выравниванию электронной населенности на этих двух уровнях, сохраняются фотоны возбуждающего излучения на частоте $\nu$. В условиях насыщения (выравнивания населенностей двухуровневого перехода) [5] эти фотоны в области частот $\nu < \nu_{01}$ распространяются со скоростью света $c = 3 \cdot 10^{10}$ см/сек. Этим фотонам, распространяющимся вдоль оси пучка накачки, в свою очередь, соответствует появление второго конуса, по образующим которого фотоны распространяются с другой величиной фазовой скорости $V = c/n(\nu)$.

Отметим то обстоятельство, что сам факт появления конусов излучения Вавилова-Черенкова свидетельствует о том, что в диспергирующей среде, показатель преломления которой больше единицы, имеет место различие в скоростях распространения фотонов. Соответствие скорости ее фазовому значению следует, например, из представленных в [1] результатов расчета угла раствора таких конусов.

Использование в эксперименте [1,2] мощных световых импульсов когерентного излучения (10- 500 кВт) длительностью 10 нсек в почти резонансных условиях позволяет выровнять населенности уровней изучаемого перехода $4S_{1/2}$ - $4P_{3/2}$ в атомарных парах калия и, тем самым, обеспечить условия «сверхсветового» (со скоростью $c$) распространения излучения на частоте $\nu$ в первую очередь на оси светового пучка.

Сам процесс выравнивания населенностей в двухуровневой среде, осуществляемый в условиях почти резонансного взаимодействия излучения накачки со средой ($\nu < \nu_{01}$), связан с трехфотонным вынужденным электронным комбинационным рассеянием (ВЭКР) [6]; наблюдаемая в этом случае [7] частота - $\nu_s$, соответствующая ВЭКР, может быть определена из соотношения:

$$\nu_s = 2\nu - \nu_{01}, \quad (1)$$

где $\nu$ - частота излучения накачки, $\nu_{01}$ – боровская частота исследуемого электронного перехода $4S_{1/2}$ - $4P_{3/2}$ в парах атомарного калия.

По мере заполнения электронами верхнего уровня $4P_{3/2}$ трехфотонный процесс ВЭКР, уравновешивается за счет ШПР, возвращающего электроны на основной, ниже-расположенный исходный уровень $4S_{1/2}$:

$$3\nu = \nu_s + 2\nu_{as}, \quad (2)$$

Взаимодействие процессов (1) и (2) не допускает создание инверсии на исследуемом переходе $4S_{1/2} - 4P_{3/2}$.

Величина частоты $\nu_{as}$, согласно [1], соответствует величине:

$$\nu_{as} = (\nu_{01} + \nu)/2. \quad (3)$$

Соотношение (2) после умножении на постоянную Планка $h$ соответствует закону сохранения энергии для ШПР в случае почти резонансного взаимодействия лазерного излучения с двухуровневой средой с разрешенным переходом.

Вероятность нелинейных процессов ВЭКР и ШПР в резонансных условиях оказывается достаточной для того, чтобы обеспечить на оси светового пучка упомянутое выше выравнивание населенностей исследуемого электронного перехода. Как правило, в случае отсутствия точного резонанса между частотой возбуждающего излучения ν и боровской частотой [8] перехода $\nu_{01}$, именно процессы ВЭКР и ШПР определяют в области спектра $\nu < \nu_{01}$ спектральную и угловую структуру спектра излучения на выходе кюветы с атомарной средой [1,2,7]. Анализ экспериментальных результатов, выполненный для случая атомарных паров калия [7], свидетельствует о том, что в почти резонансных условиях в случае использования импульсной накачки процессы низшего порядка – не эффективны и не вносят заметного вклада в наблюдаемую картину взаимодействия излучения со средой.

Таким образом, нелинейные процессы ВЭКР и ШПР в атомарных парах калия приводят к формированию двух конусов черенковского излучения.

Подобные конусы наблюдались в работе [9], где меньшая величина частотной расстройки между линиями главного дублета в атомарном натрии, чем в калии, позволяет сделать предположение, что обе линии дублета в случае натрия могут вносить вклад в конусную структуру выходящего из кюветы излучения, что приводит к усложнению ее формы. В работе [9] содержится список работ по наблюдению конусного излучения другими группами исследователей.

Для оценки угла раствора $\Theta$ черенковского конуса, появление которого обусловлено излучением на частоте накачки ν на участках спектра $\nu < \nu_{01}$ можно воспользоваться соотношением:

$$\cos \Theta = 1 / n(\nu). \qquad (4)$$

Мы подробнее остановимся на объяснении причины замедления фотонов в области тени вдоль конусных образующих. Их распространение со скоростью $V = c / n(\nu)$ приводит к формированию одного из конусов черенковского излучения. Действительно, если используются одномодовые лазерные пучки с гауссовым распределением интенсивности по сечению пучка, то при их распространении сквозь исследуемую среду в случае $\nu < \nu_{01}$ именно на оси пучка, в первую очередь, можно осуществить просветление среды (атомарных паров калия с низкочастотной стороны электронного перехода $4S_{1/2} - 4P_{3/2}$), т.е. согласно соотношениям (1) и (2) обеспечить равенство населенностей на указанных уровнях и тем самым, именно, на оси пучка создать условия для распространения части фотонов пучка излучения накачки со скоростью света $c$. Фактически, в этом случае условия для появления излучения Вавилова-Черенкова – выполнены.

То обстоятельство, что на периферии сечения пучка в области «тени» в соответствии с теорией дисперсии [4] на вышеуказанном участке спектра в атомарных парах калия показатель преломления больше единицы, и способствует появлению конуса черенковского излучения, что явным образом свидетельствует о распространении фотонов по образующим конуса с фазовой скоростью $V = c / n(\nu)$.

Согласно принципа соответствия Н. Бора [10] для обсуждения результатов эксперимента допустимо одновременное использования, с одной стороны, результатов классической теории, в первую очередь, выводов классической теории дисперсии. С другой стороны - квантовому представлению светового излучения отвечает использование представлений о корпускулярных свойствах светового излучения.

Н. Бор в работе [10] сформулировал вторую часть принципа соответствия таким образом: «между однозначным применением понятия стационарных состояний и механическим анализом внутриатомных движений существует то же соотношение дополнительности, какое существует между световым квантом и электромагнитной теорией излучения».

Исследуемому в атомарных парах калия электронному переходу между упомянутыми в принципе соответствия Н. Бора стационарными состояниями $4S_{1/2}$ и $4P_{3/2}$ соответствует боровская частота, характеризующая возможную собственную частоту колебаний оптического электрона $v_{01}$. В соответствии с принципом дополнительности Н. Бора именно с величиной этой частоты, отвечающей «внутриатомным движениям» связаны выводы классической теории дисперсии [4], что позволяет в области частот, меньших резонансной (там, где $n(v) > 1$), объяснить замедленную скорость распространения фотонов в диспергирующей среде, благодаря чему и формируются конус излучения Вавилова-Черенкова.

### 3. Заключение

Экспериментальные результаты по наблюдению конусного излучения свидетельствуют о том, что в диспергирующей среде распространение фотонов вдоль конусных образующих с разными скоростями является естественным следствием зависимости показателя преломления среды от частоты. При этом следует отметить, что скорость распространения фотонов вдоль конусных образующих, соответствующая фазовой скорости света, практически, не зависит от изменений показателя преломления среды на оси светового пучка, вызываемых мощным излучением накачки, так как эти образующие, фактически, формируются в области тени и не перекрываются с осью пучка излучения накачки – областью распространения источников, вызывающих черенковское излучение.

Отметим, что подход, учитывающий взаимодействия излучения с двухуровневыми средами в отсутствии точного резонанса, представленный в настоящей работе, также может быть использован для интерпретации некоторых экспериментальных результатов по фотолюминесценции [11], обычно запаздывающей относительно возбуждающего излучения, рэлеевскому рассеянию света в различных средах [12].

### 4. Список литературы

# On speed of a photon in a dispersing medium

Ogluzdin Valeriy E.

## Abstract

Based on the author of the experimental results and their treatment, a model that demonstrates that the real dispersive medium in spectral regions where the refractive index greater than unity, the velocity of propagation of photons in this environment corresponds to the phase velocity of light. These portions of the spectrum (for example, atomic potassium vapor or other alkali metal) are from low-frequency side of the main lines of the doublet and are available for research using tunable lasers.
In the calculations took into account the fact that the atomic vapor of alkali metal used as a dispersing medium, in accordance with the theory of dispersion in the above sections of the spectrum, on the periphery of the cross section of laser beam with Gaussian intensity distribution across the beam - in the shadows refractive index $n(v)$ is greater than unity, and the axis of the beam, where the radiation intensity is maximum - for radiation following the leading edge of the light pulse, the refractive index is unity, which ensures smooth on the beam axis, without any delay "superluminal "propagation of light with velocity $c$. This fact causes the appearance of the shadow cone of Vavilov - Cherenkov radiation, indicating a slow propagation of photons from the source of the cone with phase velocity $c/n(v)$.

Valeriy E. Ogluzdin
Sovjet Union, 119991,Moscow , 38 Vavilova st.
Prokhorovs General Physics Institute.

Контактный тел. (499) 503 81 98
Факс: (499)135 81 91
ogluzdin@kapella.gpi.ru